\documentclass[sigconf]{acmart}
\setcopyright{none}
\AtBeginDocument{%
  \providecommand\BibTeX{{%
    \normalfont B\kern-0.5em{\scshape i\kern-0.25em b}\kern-0.8em\TeX}}}

\usepackage{xcolor} 
\usepackage{multirow}
\usepackage{hhline}
\usepackage{pgfplots}
\usepackage{subcaption}
\usepackage{float}
\usepackage{soul}
\pgfplotsset{compat=1.18}
\usepackage{scalerel,xparse}
\NewDocumentCommand\sparkemoji{}{\scalerel*{\includegraphics{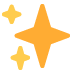}}{X}}
\usepackage{balance}

\definecolor{mycolor1}{RGB}{144, 177, 219} 
\definecolor{mycolor2}{RGB}{178, 164, 216}
\definecolor{mycolor3}{RGB}{206, 154, 156}
\definecolor{mycolor4}{RGB}{195, 216, 159}

\definecolor{mycolor5}{HTML}{4878CF} 
\definecolor{mycolor6}{HTML}{EE854A} 
\definecolor{mycolor7}{HTML}{6ACC64} 
\definecolor{mycolor8}{HTML}{D65F5F} 

\copyrightyear{2023}
\acmYear{2023}
\setcopyright{acmlicensed}\acmConference[RecSys '23]{Seventeenth ACM Conference on Recommender Systems}{September 18--22, 2023}{Singapore, Singapore}
\acmBooktitle{Seventeenth ACM Conference on Recommender Systems (RecSys '23), September 18--22, 2023, Singapore, Singapore}
\acmPrice{15.00}
\acmDOI{10.1145/3604915.3608801}
\acmISBN{979-8-4007-0241-9/23/09}

\begin{document}

\title{\scalebox{1.4}{\sparkemoji} Going Beyond Local: Global Graph-Enhanced Personalized News Recommendations}

\author{Boming Yang}
\affiliation{%
  \institution{The University of Tokyo}
  \city{Tokyo}
  \country{Japan}
}
\email{boming@g.ecc.u-tokyo.ac.jp}

\author{Dairui Liu}
\affiliation{%
  \institution{University College Dublin}
  \city{Dublin}
  \country{Ireland}}
\email{dairui.liu@ucdconnect.ie}

\author{Toyotaro Suzumura}
\affiliation{%
  \institution{The University of Tokyo}
  \city{Tokyo}
  \country{Japan}
}
\email{suzumura@ds.itc.u-tokyo.ac.jp}

\author{Ruihai Dong}
\affiliation{%
 \institution{University College Dublin}
 \city{Dublin}
 \country{Ireland}}
\email{ruihai.dong@ucd.ie}

\author{Irene Li}
\affiliation{%
  \institution{The University of Tokyo}
  \city{Tokyo}
  \country{Japan}
}
\email{bkjl6178@g.ecc.u-tokyo.ac.jp}

\renewcommand{\shortauthors}{}

\begin{abstract}
Precisely recommending candidate news articles to users has always been a core challenge for personalized news recommendation systems. Most recent works primarily focus on using advanced natural language processing techniques to extract semantic information from rich textual data, employing content-based methods derived from local historical news. However, this approach lacks a global perspective, failing to account for users' hidden motivations and behaviors beyond semantic information. To address this challenge, we propose a novel model called \textbf{GLORY} (\textbf{G}lobal-\textbf{LO}cal news \textbf{R}ecommendation s\textbf{Y}stem), which combines global representations learned from other users with local representations to enhance personalized recommendation systems. We accomplish this by constructing a Global-aware Historical News Encoder, which includes a global news graph and employs gated graph neural networks to enrich news representations, thereby fusing historical news representations by a historical news aggregator. Similarly, we extend this approach to a Global Candidate News Encoder, utilizing a global entity graph and a candidate news aggregator to enhance candidate news representation. Evaluation results on two public news datasets demonstrate that our method outperforms existing approaches. Furthermore, our model offers more diverse recommendations\footnote{Our code is released on \url{https://github.com/tinyrolls/GLORY}}.
\end{abstract}





\maketitle

\section{INTRODUCTION}
News Recommendation (NR) is the process of recommending news articles to users to satisfy their need for information by optimizing the accuracy of predicting relevance between news articles and users. Compared with making recommendations in other domains, news recommendation is more challenging because of the highly dynamic environment caused by the natural characteristics of news articles, such as timeliness, novelty, etc., for example, rapid changes occur in the relevance of news articles \cite{NewsRSFuture}. Therefore, unsurprisingly, content-based recommendation methods \cite{NRMS, NPA, NAML, LSTUR, HieRec, MINER} have proven their effectiveness. Typically, these are achieved via harnessing various natural language processing (NLP) and machine learning (ML) technologies to extract user interest representations by analyzing news articles that they have read in the past, to build representations of candidate news articles by studying their content, and then match users with candidate news articles.

Recently, deep learning-based technologies have been developed rapidly and brought new opportunities for enhancing content-based solutions due to their strong ability to deal with textual data, such as news content, news title, etc., and capture sequential dependencies in user reading history. For example, \cite{NRMS} defines a news encoder and a user encoder based on multi-head self-attention mechanisms, which learn news representations from titles and learn user representation from news embeddings. \cite{NAML} acquires additional semantic information by using both word-level and view-level attention mechanisms to the news encoder to select important words and views for learning informative news representations. Research on modeling user preferences has gained popularity. \cite{HieRec} employs a three-layer hierarchical structure to learn user interests from the subtopic and topic levels of news articles. Additionally, graph-based methods represent another direction in this field. For instance, \cite{DIGAT} utilizes a semantic augmentation graph to enhance news items and leverages a dual-graph interaction method to learn both candidate news and user representations.

\begin{figure*}[htbp]
\centering
\includegraphics[width=0.7\textwidth]{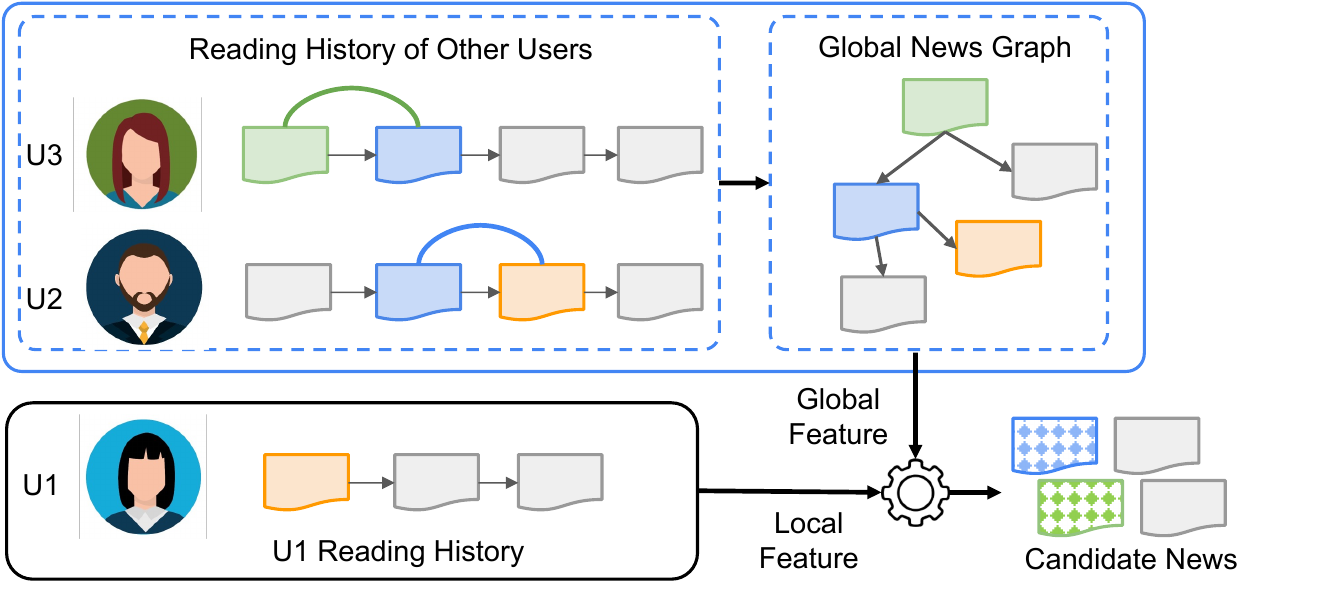}
\caption{An example of how global feature may help predict the candidate news. The figure depicts three users, U1, U2, and U3, along with three distinct historical news articles represented by green, blue, and orange colors. Dotted blue and green candidate news articles are related to the content of the blue and green news articles, respectively.}
\label{fig: intro}
\end{figure*}

Whilst these solutions have achieved great success in news recommendation tasks, such as alleviating the serious item user behaviors sparse problem by extracting meaning representation from news content, these methods mainly focus on the reading history of each user individually (i.e., using a single user's historical news click sequence to construct this user's representation), which lack a global border view of news articles and clicking among multiple users, and might be insufficient for uncovering more implicit hidden user behaviors. For example, as illustrated in Fig.~\ref{fig: intro}, assume that we want to recommend news articles for user U1. User U1's reading list contains only an orange news article, but at the moment, there are no news articles in the candidate set similar to the orange one. If recommendation models merely consider U1's reading history, it is difficult to make recommendations for U1 in this situation. But if we can harness multiple user histories together, for example, extracting a Global News Graph from user U3 and U2's clicks, we may find that there are certain relationships between orange news articles with blue news articles and green news articles. Therefore, the model may consider recommending dotted blue and green candidates that are similar to the blue one and the green one respectively. The challenge is how to properly generate this global news graph based on clicking from multiple users and integrate it with news recommendation systems.

To address the mentioned issue, we propose a novel model,  \textbf{GLORY} (\textbf{G}lobal-\textbf{LO}cal news \textbf{R}ecommendation s\textbf{Y}stem). 
Because historical news interaction data can provide more extensive and implicit relational information than semantic relevance, we propose a global-aware historical news encoder, specifically, using a global news graph to provide information on global perspectives for historical news. At the same time, to address the user behaviors sparse problem of candidate news, we propose a global-aware candidate news encoder, which uses a global entity graph to provide more effective associations for candidate news. Next, we use the multi-head self-attention mechanism to extract user interests from historical news. The final matching score came from the user news vector and the candidate news vector. We conducted extensive experiments on the MIND news dataset and Adressa, and the results showed that our model outperformed existing methods.

The main contributions of this work can be summarized as follows: 1) To the best of our knowledge, we are the first to propose a global perspective for constructing a homogeneous global news/entity graph in the news recommendation domain, which enables more effective utilization of rich historical interaction information; 2) We introduce a global-aware historical news encoder and a global-aware candidate news encoder that leverage the global news graph and global entity graph, respectively, to enhance the representations of historical news and candidate news; 3) Extensive experiments on real-world datasets demonstrate that GLORY achieves state-of-the-art performance. 

\section{RELATED WORK}

\subsection{Feature-based News Recommendation}
Traditional recommendation methods leverage features of users or news from the reading history. Typically, the factorization machine method is utilized to learn information from interaction data. For instance, LibFM \cite{LibFM} employs factorization machines (FM) to obtain features of each news, and then uses the learned features to conduct recommendations. DeepFM \cite{DeepFM}, on the other hand, utilizes an FM component and a deep component to capture more profound connection information. However, the limitation of these approaches lies in the fact that candidate news items often have sparse or no interaction history, making it challenging to learn effective representations.

\begin{figure*}[htbp]
    \centering
    \includegraphics[width=0.8\textwidth]{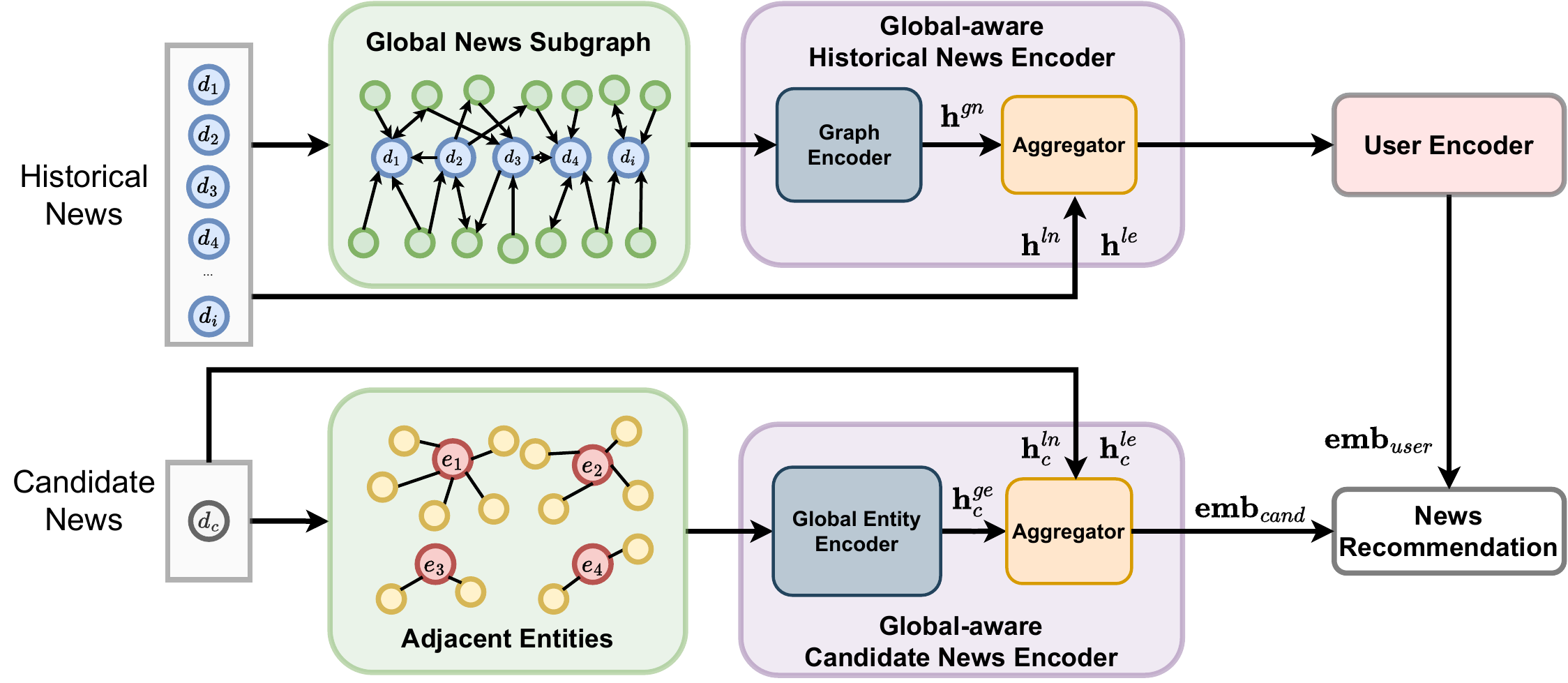}
    \caption{An overview of the proposed GLORY model. The blue node $d_i$ represents historical news in a user's behavior, while the grey node $d_c$ signifies candidate news $d_c$. The red node $e_i$ in the adjacent entities denotes an entity belonging to $d_c$. The embedding $\mathbf{h}^{ln}$ and $\mathbf{h}^{ln}_c$ denote the local news representations, and the embedding $\mathbf{h}^{le}$ and $\mathbf{h}^{le}_c$ denote the local entity representations. $\mathbf{emb}_{user}$ and $\mathbf{emb}_{cand}$ are user presentation and candidate news representation, both used for news recommendation.}
    \label{fig: main}
\end{figure*}

\subsection{Neural News Recommendation}
One advantage of news recommendation scenarios is the richness of textual information. With the rapid development of NLP techniques, research has started to focus more on content-based news recommendation systems. For example, NAML \cite{NAML} utilizes convolutional neural networks (CNNs) \cite{hughes2017medical} to learn news representations and gated recurrent units (GRUs) \cite{chung2014empirical} to learn user representations. NRMS \cite{NRMS}, on the other hand, employs multi-head self-attention mechanisms for both news and user modeling. Moreover, some models feature complex structures for news and user modeling, such as HieRec \cite{HieRec}, which uses a three-layer hierarchical structure combined with topics and subtopics to learn user representations; and MINER \cite{MINER}, which employs a poly attention scheme with disagreement regularization and category-aware attention weighting to obtain 32 user interest vectors.

\subsection{Graph-based News Recommendation}
NLP using graph-based techniques has emerged as a prominent field of research and development in recent years \cite{Gnn_NLP, graph1, graph2}. Rather than relying solely on NLP for modeling semantic information in news text, certain approaches have embraced the use of graphs to learn news representations. For example, GERL \cite{GERL} learns text-based news representations from news titles and topics, and learns graph-based news representations from the user-news graph. GNewsRec \cite{GNewsRec} utilizes a heterogeneous user-news-topic graph to discover long-term interests. User-as-graph \cite{User-as-Graph} proposes a heterogeneous graph pooling method to aggregate the local news graph formed by a user's reading history. DIGAT \cite{DIGAT} takes advantage of a semantic augmentation graph, constructed using semantically related methods, to enrich candidate news content.

Nonetheless, the mentioned related work, including content-based and graph-based methods, focuses more on complex models for contextual information to model the current user, lacking a comprehensive understanding of global news and failing to utilize the rich global user-news interaction information effectively. 
In contrast, our proposed model GLORY considers global news graph information, capturing users' latent behavioral patterns and enriching the news representations beyond merely local news information.

\section{METHOD}

Our proposed approach, GLORY, focuses on enhancing historical news representation by utilizing a global news graph, and improving candidate news representation through a global entity graph, as depicted in Fig.~\ref{fig: main}. First, we learn the representation of news text and news entities from a local perspective. Subsequently, we adopt the global-aware historical news encoder and the global-aware entity news encoder. Lastly, we employ a concise user encoder and a news recommendation component.

\subsection{Problem Formulation}
The click history sequence of a user $u$ can be denoted as $\mathbf{H}_u=[d_1, d_2, ..., d_H]$, where $H$ is the number of historical news articles. Each news article $\mathbf{d}_i$ has a title, which contains a text sequence $\mathbf{T}_i=[w_1, w_2, ..., w_T]$ consisting of $T$ word tokens, and an entity sequence, which is denoted by $\mathbf{E}_i=[e_1, e_2, ..., e_E]$ consisting of $E$ entities. The objective is to predict the level of interest $s_{u,c}$ for a given candidate news article $d_c$ and user $u$, which reflects the likelihood of a clicking action occurring between them. The recommendation task involves ranking multiple candidate news articles based on probability scores.

\subsection{Local Representation}
First, we need to learn local news representation $\mathbf{h}^{ln}$ and local entity representation $\mathbf{h}^{le}$ from news text and news entities. For each user behavior, we only consider the latest $L_{his}$ clicked news from the click history sequence $\mathbf{H_u}$ of user $u$ and candidate news $d_c$. In this work, we use only the news title as the news text, and for each news title, we truncate the first $L_{title}$ words as input.

\begin{figure*}[htbp]
    \centering
    \includegraphics[width=0.8\textwidth]{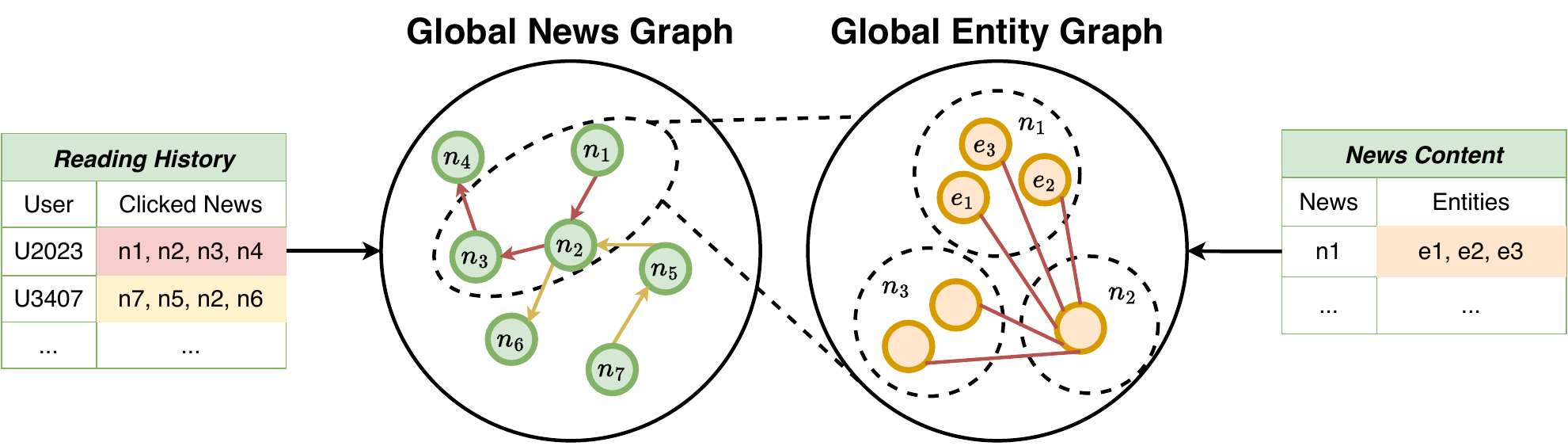}
    \caption{A simple Case of Global News Graph and Global Entity Graph. Global News Graph uses two reading sequences $[n_1, n_2, n_3, n_4]$ and $[n_7, n_5, n_2, n_6]$ from user U2023 and U3407, respectively. And Global Entity Graph shows six entities from news $n_1, n_2, n_3$, and displays the entities $e_1, e_2, e_3$ from news $n_1$.}
    \label{fig: graph}
\end{figure*}

\subsubsection{Local News Representation}
The local news encoder aims to obtain local news representation from the news text. It comprises a word embedding layer, where we use pre-trained GloVe \cite{glove} word embeddings as initialization, for converting a news text from a sequence of words $[w_1, w_2, \ldots,w_T]$ into a sequence of word embedding vectors $\mathbf{x}_n = [x_1^\omega;x_2^\omega;\ldots;x_T^\omega]$, followed by a text multi-head self-attention (MSA) \cite{attention} layer for learning word representations $\mathbf{X}_n$: 
\begin{align}
    \mathbf{X}_n = \text{MSA}(\mathbf{x_n}) &= \text{Concat}(\text{head}_1, \dots, \text{head}_h) \cdot \mathbf{W}^O, \\
    \text{where} \quad \text{head}_i &= \text{Att}(\mathbf{x}_n \cdot \mathbf{W}^Q_i, \mathbf{x}_n \cdot \mathbf{W}^K_i, \mathbf{x}_n \cdot \mathbf{W}^V_i), \\
    \text{Att}(Q, K, V) &= softmax(\frac{QK^\top}{\sqrt{d_k}})V.
\end{align}

Here, $\text{Concat}$ is the concatenation function, and $\mathbf{W}^O$ is a trainable matrix used to concatenate the outputs. For each head $i$, we use three matrix transformations $\mathbf{W}^Q_i$, $\mathbf{W}^K_i$, $\mathbf{W}^V_i$ to map the input $\mathbf{x}_n$ to query, key, and value vectors. The $\text{Att}(\cdot)$ function adopts a Scaled Dot-Product Attention to compute a weighted sum of the value vectors,  resulting in the output vector for each head. And $d_k$ denotes the dimension of keys.

Since each user has multiple news in the history, we then adopt a text attention layer to learn local news representation 
$\mathbf{h}^{ln}$ by aggregating word representations:
\begin{equation}\label{eq: attn}
    \mathbf{h}^{ln} = \sum_{i=1}^{T} \alpha_i^\omega x_i^\omega, \quad \alpha_i^\omega = \frac{exp(\mathbf{q}_\omega^\top \tanh(\textbf{W}^{\omega} \cdot \mathbf{X}_n^i))}{\sum_{j=1}^T exp(\mathbf{q_\omega^\top} \tanh(\textbf{W}^{\omega} \cdot  \mathbf{X}_n^j))},
\end{equation}

where $\textbf{W}^{\omega}$ is a trainable matrix, and $\mathbf{q}_\omega$ is the query vector. And $\mathbf{X}_n^i$ denotes the $i$-th word representation of $\mathbf{X}_n$.

\subsubsection{Local Entity Representation}
In addition, we employ a local entity encoder to learn local entity representations. The process begins with an entity embedding layer that merges information from the knowledge graph, where we use a pre-trained TransE \cite{TransE} entity embedding based on WikiData \footnote{\url{https://www.wikidata.org/wiki/Wikidata:MainPage}}, to convert a sequence of entities into a sequence of entity embedding vectors $\mathbf{x}_e = [x_1^e;x_2^e;\ldots;x_E^e]$. Subsequently, we use an entity self-attention network to learn entity representations $\mathbf{X}_e$ and an attention network to aggregate multiple entities and obtain the local entity representation $\mathbf{h}^{le}$, as in Eq ~\ref{eq: attn}.

\subsection{Global-aware Historical News Encoder}
To go beyond the local, we introduce a novel global-aware historical news encoder to merge global and local information. 

\subsubsection{Global News Graph} \label{sec: gnews}
The Global News Graph is used to summarize users' reading histories. On the left part of Fig.~\ref{fig: graph}, we show a directed Global News Graph, denoted as $G_n = (V_n, E_n)$, where $V_n$ and $E_n$ represent the sets of news articles and edges, respectively. 
To construct this graph, we collect the edges from each user's reading history in the training dataset. Specifically, a directed edge $(v_i, v_j)$ is added to the graph to indicate that news item $v_j$ was read immediately after $v_i$ by the same user. The edge weight is determined by the frequency of this occurrence from all reading history.

\subsubsection{Graph Encoder}
For each user history, we extract the subgraph $G_{sub}^u = (V_{sub}^u, E_{sub}^u)$ from the global news graph to capture a global perspective on a specific user's interests. For each news in $\mathbf{H_u}$ of user $u$, we select news neighbors of multiple hops from the global news graph. For each hop, we select the top $M_n$ neighbors based on the edge weight. We adopt graph neural networks (GNN) \cite{li2015gated} to encode the subgraph to obtain global news embedding $\mathbf{h}^{gn}$:

\begin{equation}
        \mathbf{h}^{gn}=\textrm{GNN}(\mathbf{h}^{ln}).
        \label{eq: ge}
\end{equation}

While there are multiple choices for the graph encoder, here we apply the gated graph neural network (GGNN) \cite{gated}, which employs a gated recurrent unit (GRU) \cite{GRU} to capture sequence-based hidden behavior information:  
\begin{gather}
    \mathbf{h}^{(0)} = \mathbf{h}^{ln}, \\
    \mathbf{m}_i^{(l+1) } = \sum_{j \in \mathcal{N}(i)} W_g \cdot \mathbf{h}_j^{(l) }, \\
    \mathbf{h}_i^{(l+1) } = \textrm{GRU} (\mathbf{m}_i^{(l+1) }, \mathbf{h}_i^{(l) }).
\end{gather}
We initialize $\mathbf{h}^{(0)}$ with $\mathbf{h}^{ln}$. $N_i$ represents the set of neighboring nodes for node $i$ in subgraph $G_{sub}^u$, $l$ denotes layer number, and $W_g$ is a trainable matrix. 

\subsubsection{Historical News Aggregator}
For each historical news article $d_k$, we have the local news representation $\mathbf{h}^{ln}$, the local entity representation $\mathbf{h}^{le}$ and the global title representation $\mathbf{h}^{gn}$. We now utilize an attention pooling network to learn the historical news representation $\mathbf{n}_k$ by aggregating these three representations $\mathbf{h}^k = [\mathbf{h}^{ln};\mathbf{h}^{le};\mathbf{h}^{gn}]$:
\begin{equation}
    \mathbf{n}_k = \sum_{i=1}^{3} \alpha_i^h \mathbf{h}_i^k, \quad \alpha_i^h = \frac{exp(\mathbf{q}_h^\top \tanh(\textbf{W}^h \cdot \mathbf{h}^k_i))}{\sum_{j=1}^3 exp(\mathbf{q_h^\top} \tanh(\textbf{W}^h \cdot \mathbf{h}^k_j))}.
    \label{eq:aggregator}
\end{equation}
where $\textbf{W}^h$ is a trainable matrix, and $\mathbf{q}_h$ is the query vector. And $\mathbf{h}_i^k$ denotes the $i$-th concatenated representation of $\mathbf{h}^k$.

\subsection{User Encoder}
Next, we briefly introduce the architecture of the user encoder in GLORY. Following \cite{NRMS}, after we obtain news embedding $\mathbf{n}_k$ of each news in a given user reading history $H_u$, we employ a multi-head attention layer and an attention pooling layer to finally learn the user representation $\textbf{emb}_{user}$, similar as Eq.~\ref{eq: attn}.

\subsection{Global-aware Candidate News Encoder}
As a vast number of news articles are generated daily, candidate news often lacks user interaction history. Therefore, we can not only enhance historical news using the global news graph but also improve candidate news using a similar global entity graph. Consequently, we propose a novel global-aware candidate news encoder. 

\subsubsection{Global Entity Graph} \label{sec: gentity}
In order to construct a global entity graph, we employed a similar approach to that used in the global news graph but adopted an undirected graph. Shown in Fig.~\ref{fig: graph}, the global entity graph is represented as $G_e = (V_e, E_e) $, where $V_e$ and $E_e$ are the sets of entities and edges. Entity edges are derived from each user's reading history on the training dataset. An undirected edge $(v_i, v_j)$ implies that $v_j$ is the entity of the last news article and $v_i$ is the entity of the subsequent news article. The edge weight of $(v_i, v_j) $ is determined by the count of news edge occurrences.

\subsubsection{Global Entity Encoder}
Due to the significantly smaller number of entities compared to the number of news articles, the average degree of nodes in the global entity graph is much higher. Therefore, it is more convenient and concise to use adjacent entities directly instead of employing entity subgraphs. Similarly, we select the top $M_e$ neighbor entities for each entity in candidate news $d_c$ based on edge weights and use the same entity embedding layer as the local-view entity encoder, the final embedding represented as $\mathbf{x}_c = [x_1^c, x_2^c, ..., x_{E*M_e}^c]$. 
Similarly, as in Eq.\ref{eq: attn}, we utilize a $MSA$ to learn the entity token representations $\mathbf{X}_c = \text{MSA}(\mathbf{x}_c)$ and an attention network to aggregate the adjacent entity representations $\mathbf{h}_c^{ge}$. 

\subsubsection{Candidate News Aggregator}
For each candidate news article $d_c$, we have the local news representation $\mathbf{h}_c^{ln}$, the local entity representation $\mathbf{h}_c^{le}$ and the global entity representation $\mathbf{h}_c^{ge}$. We use an attention pooling network to learn the candidate news representation $\mathbf{emb}_{cand}$ by aggregating these three representations $\mathbf{h}^c = [\mathbf{h}_c^{ln};\mathbf{h}_c^{le};\mathbf{h}_c^{ge}]$, as in Eq.~\ref{eq:aggregator}.

\subsection{News Recommendation}
In accordance with the prior work \cite{NPA}, we employ the negative sampling technique during our training process. For every behavioral session in the training dataset, each clicked candidate news is denoted as a positive sample $n_i^+$, while we randomly select $K_{neg}$ non-clicked candidate news within the same session as negative samples $[n_i^1, n_i^2, ..., n_i^{K_{neg}}]$. Subsequently, we estimate the click score for each news item as $\mathbf{\hat{y}_i} = [\hat{y}_i^+, \hat{y}_i^1, \hat{y}_i^2, ..., \hat{y}_i^{K_{neg}}]$, by conducting an inner product of user embedding $\mathbf{emb}_{user}$ and all samples $\mathbf{emb}_{cand}$, which contains one positive sample and $K_{neg}$ negative samples. Finally, we optimize the positive samples log-likelihood loss $\mathcal{L}_{NCE}$ over the training dataset during model training:
\begin{equation}
    \hat{\mathbf{y}_i} = softmax(\mathbf{emb}_{user} \cdot \mathbf{emb}_{cand}) ,
\end{equation}
\begin{equation}
    \mathcal{L}_{NCE} = - \sum_{i=1}\log\frac{exp(\hat{\mathbf{y}}_i^+) }{ exp(\hat{\mathbf{y}}_i^{+})  + \sum_{j=1}^{K_{neg}} exp(\hat{\mathbf{y}}_i^j) }.
\end{equation}

\section{Experimental Setup}
\subsection{Datasets}
We conduct extensive experiments on two real-world datasets to evaluate the performance of our model. The first one is the MIND dataset\cite{MIND}, which was collected from anonymized behavior logs of Microsoft News\footnote{\url{https://news.microsoft.com}} website. The full version of MIND (MIND-large) includes a million sampled users, who had at least 5 news clicks during 6 weeks from October 12 to November 22, 2019. In addition, a small version of MIND (MIND-small) was also released by sampling 50,000 users and their behavior logs. Another dataset is Adressa \cite{Adressa}, which was collected from Neiwegen News Website Adresseavisen\footnote{\url{https://www.adressa.no/}}, and here we use the light version which includes 1 week (from 1 January to 7 January 2017). Key statistics of datasets are shown in Tab.~\ref{tab: datasets}.

\begin{table}[htbp]
    \caption{Dataset statistics.}

    \centering
    \begin{tabular}{lrrr}
    \toprule
              & MIND-small & MIND-large & Adressa \\
    \midrule
    \# News   & 65,238     & 161,013    & 14,732           \\
    \# Users  & 50,000     & 1,000,000  & 537,627            \\
    \# Clicks & 347,727    & 24,155,470 & 2,527,571          \\
    \bottomrule
    \end{tabular}
    \label{tab: datasets}
\end{table}

\subsection{Evaluation metrics}
Following the setting from \cite{MIND}, we evaluate recommendation performance via four metrics, Area under the ROC Curve (AUC), Mean Reciprocal Rank (MRR), and normalized discounted cumulative gain (nDCG@N)  with $N = 5, 10$.

\begin{table*}[htbp]
    \caption{Evaluation performance of all baselines on MIND. There are some missing values because we reuse the results from existing works: $\ast$ is taken from\cite{MINER}; $\dagger$ is taken from\cite{DIGAT}; $\ddagger$ is taken from \cite{User-as-Graph}. }
    \centering
    \begin{tabular}{c | c c c c | c c c c}
    \toprule
    \multirow{2}{*}{\textbf{Method}} & \multicolumn{4}{c|}{\textbf{MIND-small}} & \multicolumn{4}{c}{\textbf{MIND-large}} \\
    \cline{2-9}
           & AUC & MRR & nDCG5 & nDCG10 & AUC & MRR & nDCG@5 & nDCG@10 \\
     \hline
     LibFM & 59.74 & 26.33 &27.95 & 34.29 & 61.85 & 29.45 & 31.45 & 37.13 \\
     DeepFM & 59.89 & 26.21 & 27.74 & 34.06 & 61.87 & 29.3 & 31.35 & 37.05 \\
     \hline
     DKN & 62.90 & 28.37 & 30.99 & 37.41 & 64.07 & 30.42 & 32.92 & 38.66 \\
     NPA & 64.65 & 30.01 & 33.14 & 39.47 & 65.92 & 32.07 & 34.72 & 40.37 \\
     NAML & 66.12 & 31.53 & 34.88 & 41.09 & 66.46 & 32.75 & 35.66 & 41.40 \\
     LSTUR & 65.87 & 30.78 & 35.15 & 40.15 & 67.08 & 32.36 & 35.15 & 40.93 \\
     NRMS & 65.63 & 30.96 & 34.13 & 40.52 & 67.66 & 33.25 & 36.28 & 41.98 \\
      FIM & 65.34 & 30.64 & 33.61 & 40.16 & 67.87 & 33.46 & 36.53 & 42.21 \\
     HieRec & 67.95 & 32.87 & \underline{36.36} & 42.53 & 69.03 & 33.89 & 37.08 & 43.01 \\
     MINER w/o PLM$\ast$ & \underline{68.07} & \underline{32.93} & - & \underline{42.62} & - & - & - & - \\
     \hline
     GERL & 65.27 & 30.10 & 32.93 & 39.48 & 68.10 & 33.41 & 36.34 & 42.03 \\ 
     GNewsRec & 65.54 & 30.27 & 33.29 & 39.80 & 68.15 & 33.45 & 36.43 & 42.10 \\
     User-as-Graph$\ddagger$ & - & - & - & - & \underline{69.23} & \underline{\textbf{34.14}} & \underline{37.21} & \underline{43.04} \\
     DIGAT w/o PLM$\dagger$ & 67.82 & 32.65 & 36.25 & 42.49 & - & - & - & -  \\
     \hline
     GLORY(ours) & \textbf{68.15} & \textbf{32.97} & \textbf{36.47} & \textbf{42.78} & \textbf{69.54} & 34.03 & \textbf{37.92} & \textbf{44.19} \\
     \bottomrule
    \end{tabular}
    \label{tab: mainresult}
\end{table*}

\subsection{Baselines}
We mainly select three groups of baselines to compare:

\textbf{Feature-based Methods} (1) LibFM \cite{LibFM}, a widely used recommendation method that uses factorization machines to predict user preferences for news articles based on their features; (2) DeepFM \cite{DeepFM}, a neural network-based recommendation method that combines factorization machines with deep learning to improve recommendation accuracy for news articles.

\textbf{Neural Recommendation Methods}: (1)  \textbf{DKN} \cite{DKN}, using a knowledge-aware CNN to learn news representation from both news text and knowledge entities; (2) \textbf{NPA} \cite{NPA}, using personalized attention network to learn news and user representation; (3) \textbf{NAML} \cite{NAML}, using the multi-view attention network learn representation based on different news article features; (4) \textbf{LSTUR} \cite{LSTUR}, jointly modeling users' long-term and short-term interests by a GRU network; (5) \textbf{NRMS} \cite{NRMS}, using the multi-head self-attention network to learn user and news representation; (6) \textbf{FIM} \cite{FIM}, utilizing hierarchical representations of news articles through stacked dilated convolutions to perform fine-grained interest; (7) \textbf{HieRec} \cite{HieRec}: using a three-level hierarchical structure to represent each user's diverse and multi-grained interest. (8) \textbf{MINER w/o PLM} \cite{MINER}: using a poly attention scheme with disagreement regularization to model multiple user interests. We select the version without the pre-trained language model (PLM) to compare fairly. 

\textbf{Graph-based Methods}: (1) \textbf{GERL} \cite{GERL}, using a bipartite graph and combines transformer architecture with a graph attention network to enhance news and user representations; (2) \textbf{GNewsRec} \cite{GNewsRec}, utilizing a heterogeneous graph to model user-news-topic interactions, and employs graph neural networks to learn high-order user and news representations; (3) \textbf{User-as-Graph} \cite{User-as-Graph}, using a personalized heterogeneous graph built from user behaviors and a novel heterogeneous graph pooling method; (4) \textbf{DIGAT} \cite{DIGAT}, using news- and user-graph channels and a dual-graph interaction process to match news-user representations effectively. Similarly, we choose the version without PLM. 

\subsection{Experiment Settings}
Following the previous work \cite{NPA}, we set the most recently clicked news article of a user as $L_{his} = 50$, and the maximum title length as $L_{title}=30$. The number of maximum used entities of each news is set to be $L_{entity} = 5$, and the number of neighboring news $M_n$, and hops $K$ in the news sub-graph are 8 and 2, respectively. The number of neighboring entities $M_e$ in adjacency entities is 10. We set the number of GGNN layers as 3. We adopt 300-dimensional pre-trained GloVe word embeddings and 100-dimensional pre-trained TransE entity embeddings for MIND datasets initialization. And the historical and candidate news representations are 400-dimensional. As for the optimizer, we choose Adam \cite{Adam} as our optimization algorithm, with a learning rate $2e^{-4}$ and linearly decayed with 10\% warm-up steps. And the negative sampling rate $K_{neg}$ is set to 4. All hyper-parameters were selected according to the result of the validation set. For the Adressa dataset, the difference is that we only use news titles, with randomly initialized word embeddings and a learning rate of $1e^{-5}$.

\section{RESULTS}

\subsection{Main Result}
Tab.~\ref{tab: mainresult} shows the performance comparison between GLORY and the baselines on two settings of the MIND dataset. The tabulated values are expressed as percentages without the symbol $\%$. The overall best and baseline best results are boldfaced and underlined respectively.  

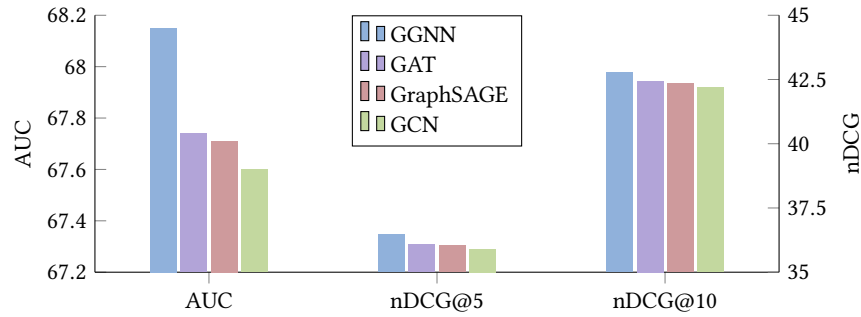
\begin{figure*}[htbp]
    \centering
    \begin{tikzpicture}
        \begin{axis}[
                ybar,
                width=0.6\linewidth,
                height=5cm,
                axis line style={-},
                axis x line*=bottom,
                axis y line*=left,
                ylabel={AUC},
                ytick distance=0.2,
                ymin=67.200, ymax=68.20,
                xtick= {4,8,12},
                xticklabels={AUC, nDCG@5, nDCG@10},
                xticklabel style={align=center},
                xmin=2, xmax=14,
                legend style={at={(0.5,1)}, anchor=north, legend cell align=left},
        ]
        \addplot[fill=mycolor1, bar shift=-0.6cm,draw=none] coordinates {(4, 68.15)};
        \addplot[fill=mycolor2, bar shift=-0.2cm,draw=none] coordinates {(4, 67.74)};
        \addplot[fill=mycolor3, bar shift=0.2cm,draw=none] coordinates {(4, 67.71)};
        \addplot[fill=mycolor4, bar shift=0.6cm,draw=none] coordinates {(4, 67.60)};
        \addlegendentry{GGNN}
        \addlegendentry{GAT}
        \addlegendentry{GraphSAGE}
        \addlegendentry{GCN}
        \end{axis}
        
        \begin{axis}[
                ybar,
                width=0.6\linewidth,
                height=5cm,
                axis line style={-},
                axis x line*=bottom,
                hide x axis,
                axis y line*=right,
                ylabel={nDCG},
                ytick distance=2.5,
                ymin=35.00, ymax=45.00,
                xtick= {4,8,12},
                xticklabels={AUC, nDCG@5, nDCG@10},
                xmin=2, xmax=14,
        ]
        \addplot[fill=mycolor1, bar shift=-0.6cm, draw=none] coordinates {(8, 36.47)};
        \addplot[fill=mycolor2, bar shift=-0.2cm,draw=none] coordinates {(8, 36.07)};
        \addplot[fill=mycolor3, bar shift=0.2cm,draw=none] coordinates {(8, 36.04)};
        \addplot[fill=mycolor4, bar shift=0.6cm,draw=none] coordinates {(8, 35.91)};

        \addplot[fill=mycolor1, bar shift=-0.6cm,draw=none] coordinates {(12, 42.78)};
        \addplot[fill=mycolor2, bar shift=-0.2cm,draw=none] coordinates {(12, 42.44)};
        \addplot[fill=mycolor3, bar shift=0.2cm,draw=none] coordinates {(12, 42.35)};
        \addplot[fill=mycolor4, bar shift=0.6cm,draw=none] coordinates {(12, 42.18)};

        \end{axis}                
    \end{tikzpicture}
    \caption{Comparison of graph encoders, results on MIND-small.}
    \label{fig: gcn}
\end{figure*}

\textbf{MIND} We have several reservations from Tab.~\ref{tab: mainresult}: 1)  The outcomes of neural news recommendation methods surpass those of feature-based methods across all cases, thereby highlighting the efficacy of neural networks in modeling news and users based on implicit semantic representation rather than manually created features. 2) While some baseline methods apply news topics and categories as auxiliary information to improve news encoding (i.e., HieRec), or design complex architecture to encode users (i.e., MINER w/o PLM constructs 32 interest vectors using a polyhedral interest system for users), they focus on a local perspective of the historical news of a given user. Our GLORY model integrates global features and outperforms these baselines. 
Although DIGAT w/o PLM employs a semantically enhanced graph to enable the model to learn beyond historical news, it is still based on semantic relevance and does not learn latent behavioral information from other users' interactions. In contrast, GLORY, even with simply attention-based user interest encoding, can achieve the best performance, and the global graph information provides significant assistance in learning news encoding. 3)  
Both GERL and GNewsRec employ graph-based methods to model user-news interactions. However, the bipartite interaction-based approach faces difficulties in the test dataset, particularly when candidate news items are novel and lack interaction history. In contrast, our approach enhances historical news by incorporating information from the global news graph, thereby augmenting user interests. We also leverage a global entity graph to strengthen the relevance of candidate news beyond just semantics.

\begin{table}[htbp]
    \centering
    \caption{Evaluation performance of several baselines on Adressa dataset. All methods only utilize the news titles.}

    \begin{tabular}{c|c|c|c|c}
    \hline
    \multirow{2}{*}{\textbf{Method}} & \multicolumn{4}{c}{\textbf{Adressa}} \\
    \cline{2-5}
           & AUC & MRR & nDCG@5 & nDCG@10 \\
     \hline
     NPA & 71.89 & 38.62 & 	38.76 & 45.72 \\
     LSTUR & 73.17 & 39.47 & 39.79 &	47.21 \\
     NAML & 73.14 & 38.10 & 38.56 & 46.13 \\
     NRMS & 72.33 & 41.08 & 41.92 & 48.57 \\
     GLORY & \textbf{74.31} & \textbf{42.29} & 
     \textbf{43.37} & \textbf{49.51}\\
     \hline
    \end{tabular}
    \label{tab: adressaresult}
\end{table}

\textbf{Adressa} Tab.~\ref{tab: adressaresult} demonstrates the performance of GLORY and several baselines on the Adressa dataset. Since we only use news titles in Adressa, without categories and entities, we select several open-source methods that rely solely on titles. We observe that even without explicit entity information, GLORY achieves the best performance due to the enhancement provided by the global news graph. This suggests that historical news interactions from other users can offer additional hidden behavioral information, which helps uncover relevance beyond semantics. LSTUR outperforms NPA, showing that the GRU method has a significant ability to detect the sequence of reading history. Furthermore, NRMS achieves the second-best MRR, nDCG@5, and nDCG@10 performances, indicating that multi-head self-attention effectively selects important news to learn more informative news and user representations. Our GLORY model takes advantage of both aspects by learning news representation through multi-head self-attention and capturing sequence information from the global news graph. Consequently, we use the same dataset content but better utilize the information to achieve improved results.

\subsection{Ablation Study}

We aim to enhance our understanding of the effectiveness of each component in GLORY by conducting ablation studies on degraded models. 
Specifically, we conduct four experiments on entity and news graph: 1)  \textbf{full model}; 2)  \textbf{w/o g-news}: without global news graph enhanced representation; 3)  \textbf{w/o g-entity}: without global entity graph enhanced representation; 4)  \textbf{w/o g-news/entity}: without both aforementioned components. In each experiment, all other settings remained unchanged as in the full model, with only one component being either replaced or removed. We evaluate those experiments on MIND-small. The results of these experiments are presented in Tab.~\ref{tab: ablation}.

\begin{table}[htbp]
    \caption{Effects of different GLORY components, result on MIND-small.}
    \centering
    \begin{tabular}{c|c c c c}
    \hline
        & AUC & MRR & nDCG@5 & nDCG@10 \\
      \hline
       full model & \textbf{68.15} & \textbf{32.97} & \textbf{36.47} & \textbf{42.78} \\
       w/o g-news & 67.53 & 32.39 & 35.89 &  42.10 \\
       w/o g-entity & 67.77 & 32.57 & 36.09 & 42.38  \\
       w/o g-news/entity & 67.28 & 32.00 & 35.38 & 41.77 \\
       \hline
    \end{tabular}
    \label{tab: ablation}
\end{table}

The \textbf{full model} performs the best, but the individual components also play a critical role in achieving optimal results. The results showed that the global news graph has the most significant impact. For instance, without the global news graph, the AUC metric dropped from 68.15 to 67.53, demonstrating the effectiveness of these graph components in enhancing news recommendations. These global graphs provide hidden information beyond local semantics and can effectively improve the performance of news recommendations.

\textbf{Graph Encoder Variations}
We then study various graph encoders Eq ~\ref{eq: ge} in this section. We compare three popular graph models with GGNN in GLORY: Graph Convolutional Network (GCN) \cite{GCN}, Graph Attention Network (GAT) \cite{GAT}, and GraphSAGE \cite{GraphSAGE}. In Fig.~\ref{fig: gcn}, we report their performance on the MIND-small. We observe that GGNN significantly outperforms the other graph models, as the GGNN utilizing GRU is capable of learning the hidden sequential information underlying users' click behavior.

\subsection{Analysis on Hyperparameters}

\begin{figure}[htbp]
    \centering
    \begin{subfigure}[b]{0.45\textwidth}
        \centering
        \begin{tikzpicture}[baseline]
        \centering
            \begin{axis}[
                width=\linewidth,
                height=5cm,
                axis line style={-},
                axis x line*=bottom,
                axis y line*=left,
                ylabel={AUC},
                ylabel style={yshift=-0.1cm, font=\small},
                ytick distance=0.2,
                ymin=67.200, ymax=68.20,
                xtick= {4,8,12,16, 20},
                xticklabels={4, 8, 12, 16, 20},
                xticklabel style={align=center},
                xlabel={$M_n$},
                xmin=2, xmax=22,
                every axis plot/.append style={line width=2pt}
            ]
            \addplot [mark=*, color=mycolor5] coordinates {(4, 67.77) (8, 68.15) (12, 67.72) (16, 68.08)(20, 67.68)};
            \end{axis}
        
            \begin{axis}[
                width=\linewidth,
                height=5cm,
                axis line style={-},
                axis x line*=bottom,
                hide x axis,
                axis y line*=right,
                ylabel={nDCG},
                ylabel style={yshift=0.3cm, font=\small},
                ytick distance=2.5,
                ymin=35.00, ymax=45.00,
                xtick=data,
                xtick= {4,8,12,16,20},
                xmin=2, xmax=22,
                legend style={
                at={(0.5,0.4)}, 
                font=\footnotesize,
                anchor=north, legend cell align=left,legend columns=-1},
                every axis plot/.append style={line width=2pt}
            ]
            \addplot [mark=square*, color=mycolor6] coordinates {(4, 35.69) (8, 36.47) (12, 35.78) (16, 36.41)(20, 36.11)};
            \addlegendentry{nDCG@5}
            
            \addplot [mark=triangle*, color=mycolor7] coordinates {(4, 42.10) (8, 42.78) (12, 42.12) (16, 42.63)(20, 42.34)};
            \addlegendentry{nDCG@10}
        
            \addlegendimage{mark=*, color=mycolor5}
            \addlegendentry{AUC}       
            \end{axis}
        \end{tikzpicture}       
        \caption{Impact of news neighbors $M_n$ (on fixed $K=2, M_e=10$).}
        \label{fig: subgraph1}
    \end{subfigure}
    \hspace{0.05\textwidth}
    \begin{subfigure}[b]{0.45\textwidth}
        \centering
        \begin{tikzpicture}[baseline]
            \centering
            \begin{axis}[
                width=\linewidth,
                height=5cm,
                axis line style={-},
                axis x line*=bottom,
                axis y line*=left,
                ylabel={AUC},
                ylabel style={yshift=-0.1cm, font=\small},
                ytick distance=0.2,
                ymin=67.200, ymax=68.20,
                xtick= {5,10,15,20},
                xticklabels={5,10,15,20},
                xticklabel style={align=center},
                xlabel={$M_e$},
                xmin=2, xmax=22,
                every axis plot/.append style={line width=2pt}
            ]
            \addplot [mark=*, color=mycolor5] coordinates {(5, 67.83) (10, 68.15) (15, 67.81) (20, 67.75)};
            \end{axis}
        
            \begin{axis}[
                width=\linewidth,
                height=5cm,
                axis line style={-},
                axis x line*=bottom,
                hide x axis,
                axis y line*=right,
                ylabel={nDCG},
                ylabel style={yshift=0.3cm, font=\small},
                ytick distance=2.5,
                ymin=35.00, ymax=45.00,
                xtick=data,
                xtick= {5,10,15,20},
                xmin=2, xmax=22,
                legend style={
                at={(0.5,0.4)}, 
                font=\footnotesize,
                anchor=north, legend cell align=left,legend columns=-1},
                every axis plot/.append style={line width=2pt}
            ]
            \addplot [mark=square*, color=mycolor6] coordinates {(5, 36.20) (10, 36.47) (15, 35.89) (20, 35.93)};
            \addlegendentry{nDCG@5}
            
            \addplot [mark=triangle*, color=mycolor7] coordinates {(5, 42.27) (10, 42.78) (15,  42.24) (20, 42.23)};
            \addlegendentry{nDCG@10}
        
            \addlegendimage{mark=*, color=mycolor5}
            \addlegendentry{AUC}       
            \end{axis}
        \end{tikzpicture}
        \caption{Impact of adjacent entities $M_e$ (on fixed $K=2, M_n=8$).}
        \label{fig: subgraph2}
    \end{subfigure}

    \caption{Recommendation performance of GLORY with different $M_n$ and $M_e$ settings.}
    \label{fig: hyper}
    
\end{figure}
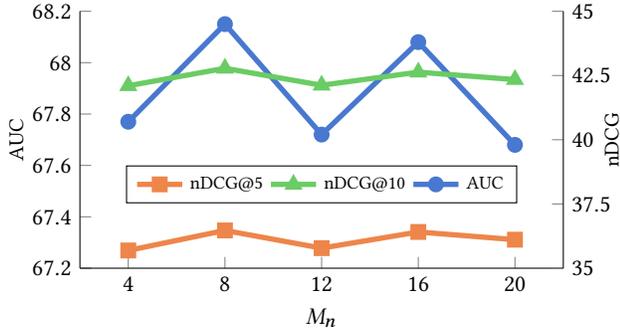
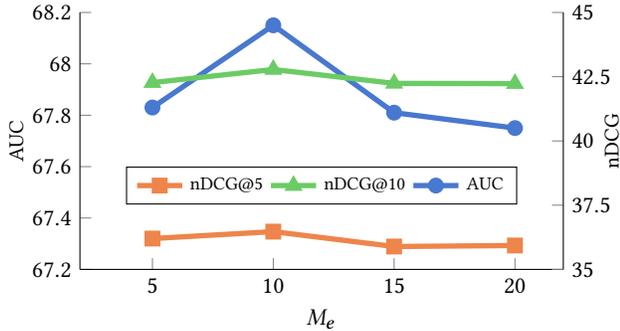

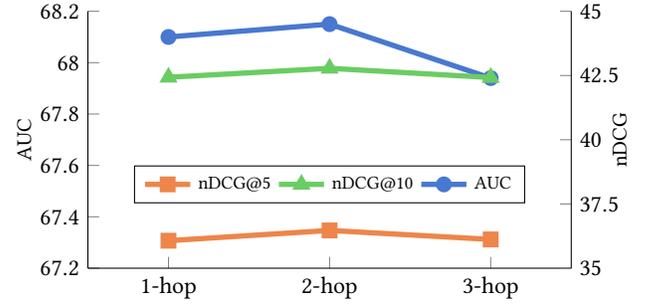
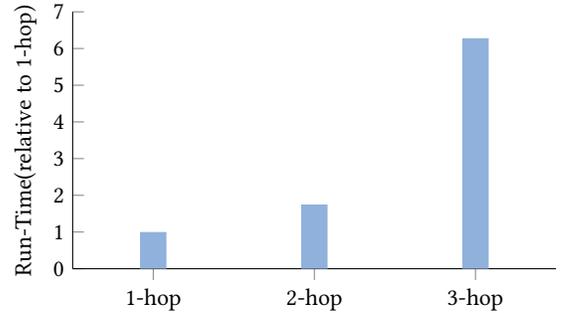
\begin{figure}[htbp]
    \centering
    \begin{subfigure}[b]{0.45\textwidth}
        \centering
        \begin{tikzpicture}
            \begin{axis}[
                width=\linewidth,
                height=5cm,
                axis line style={-},
                axis x line*=bottom,
                axis y line*=left,
                ylabel={AUC},
                ylabel style={yshift=-0.1cm, font=\small},
                ytick distance=0.2,
                ymin=67.200, ymax=68.20,
                xtick= {2,4,6},
                xticklabels={1-hop, 2-hop, 3-hop},
                xticklabel style={align=center},
                xmin=1, xmax=7,
                every axis plot/.append style={line width=2pt}
            ]
            \addplot [mark=*, color=mycolor5] coordinates {(2, 68.10) (4, 68.15) (6, 67.94)};
            \end{axis}

            \begin{axis}[
                width=\linewidth,
                height=5cm,
                axis line style={-},
                axis x line*=bottom,
                hide x axis,
                axis y line*=right,
                ylabel={nDCG},
                ylabel style={yshift=0.3cm, font=\small},
                ytick distance=2.5,
                ymin=35.00, ymax=45.00,
                xtick= {2,4,6},
                xticklabels={1-hop, 2-hop, 3-hop},
                xmin=1, xmax=7,
                legend style={
                at={(0.5,0.4)}, 
                font=\footnotesize,
                anchor=north, legend cell align=left,legend columns=-1},
                every axis plot/.append style={line width=2pt}
            ]
            \addplot [mark=square*, color=mycolor6] coordinates {(2, 36.07) (4, 36.47) (6, 36.12)};
            \addlegendentry{nDCG@5}
            
            \addplot [mark=triangle*, color=mycolor7] coordinates {(2, 42.43) (4, 42.78) (6,   42.42)};
            \addlegendentry{nDCG@10}
        
            \addlegendimage{mark=*, color=mycolor5}
            \addlegendentry{AUC}       
            \end{axis}
        \end{tikzpicture}
        \caption{Impact of different hops ($K$) (on fixed $M_e=10, M_n=8$).}
    \end{subfigure}
    \hspace{0.05\textwidth}
    \begin{subfigure}[b]{0.45\textwidth}
        \centering
        \begin{tikzpicture}
            \begin{axis}[
                ybar,
                width=\linewidth,
                height=5cm,
                axis line style={-},
                axis x line*=bottom,
                axis y line*=left,
                ylabel={Run-Time(relative to 1-hop)},
                ytick distance=1,
                ymin=0, ymax=7,
                xtick= {2,4,6},
                xticklabels={1-hop, 2-hop, 3-hop},
                xticklabel style={align=center},
                xmin=1, xmax=7,
            ]
            \addplot [fill=mycolor1,draw=none] coordinates {(2, 1.00) (4, 1.75) (6, 6.28)};
            \end{axis}
        \end{tikzpicture}
        \caption{Impact of different hops ($K$) (on fixed $M_e=10, M_n=8$).}
    \end{subfigure}
    \caption{Performance of GLORY with different $K$ settings.}
    \label{fig: hops}
\end{figure}

We investigate three key hyperparameters, i.e., the number of news neighbors $M_n$ and hops $K$, and the number of adjacent entities $M_e$. We plot the results on MIND-small in Fig.~\ref{fig: hyper} to show the effect of $M_n$, $K$, and $M_e$ settings. 
As shown in Fig.~\ref{fig: subgraph1}, with a fixed $K=2$ and $M_n=8$, the performance of GLORY is optimal when $M_e=10$, suggesting that adjacent entities indeed help uncover more related information, but an excessive number of such entities introduce too much noise, thereby reducing the model's performance. Similarly, in Fig.~\ref{fig: subgraph2}, when $K=2$ and $M_e=10$ are fixed, GLORY performs best with $M_n=8$, revealing that the model's performance does not continuously improve with an increasing number of news node neighbors while extracting global news subgraphs. This may be because having too many news neighbors exceeds the range of effective associations, leading to weakly related news interfering with the accurate information in the subgraph. Additionally, we test the model's performance using various values of $K$. The findings presented in Fig.~\ref{fig: hops} demonstrate that a depth of $K=2$ is optimal; Subgraphs with insufficient depth fail to encompass the entirety of the reading path, whereas overly deep subgraphs introduce excessive noise due to their larger scale and consume excessive time.

\subsection{Impact of Graph Construction}

\begin{table*}[htbp]
    \centering
    \caption{Effects of different Graph Construction on Global Entity Graph. The edge num. indicates the number of directed edges, and an undirected edge is counted as two directed edges.}
    \begin{tabular}{c|c |c c c c} 
    \hline
        Method & edge num. & AUC & MRR & nDCG@5 & nDCG10 \\
        \hline
      Intra-news \& Inter-news Directed & 522,595  &  67.86 & 32.67 & 36.13 & 42.42\\
      Intra-news \& Inter-news Undirected & 870,362  & 67.73 & 32.59 & 36.13 & 42.40\\
      Inter-news Directed & 506,507 & 67.75 & 32.59 & 36.04 & 42.35 \\
      Inter-news Undirected & 841,722 & \textbf{68.15} & \textbf{32.97} & \textbf{36.47} & \textbf{42.78} \\
      \hline
    \end{tabular}
    \label{tab: gentity}
\end{table*}

\subsubsection{Global News Graph}
We investigate the effects of different global news graph construction methods as described in Sec ~\ref{sec: gnews}, particularly focusing on the connections between the edges of different news articles. 

We conducted three experiments to analyze the effectiveness of the global news graph construction. These experiments are as follows:
1) \textbf{Dir-Seq (Directed Sequential)}: In this experiment, the historical news of each user is treated as a directed sequence, in which the preceding and succeeding news articles are connected in chronological order; 2) \textbf{Undir-Seq (Undirected Sequential)}: This experiment is based on the previous experiment, with all directed edges transformed into undirected edges; 3) \textbf{(Co-occur) Co-occurrence}: In this experiment, undirected edges are created between every pair of news articles that appear in the same history sequence. We show the results on MIND-small in Tab.~\ref{tab: gnews}.

\begin{table}[htbp]
    \caption{Effects of different Graph Construction on Global News Graph.}
    \centering
    \begin{tabular}{c |c c c c}
    \hline
       Methods & AUC & MRR & nDCG@5 & nDCG@10  \\
     \hline
       Dir-Seq & \textbf{68.15} & \textbf{32.97} & \textbf{36.47} & \textbf{42.78} \\
       Undir-Seq  & 67.73 & 32.64 & 36.07 & 42.38\\
       Co-oc  & 67.76 & 32.47 & 35.95 & 42.23 \\
       \hline
    \end{tabular}
    \label{tab: gnews}
\end{table}

The experimental results demonstrate that the directed sequential method surpasses the other two methods in performance. The undirected graphs, which include the undirected sequential and co-occurrence compositions, are unable to utilize the user's reading sequence effectively. Furthermore, the undirected sequential method performs better than the co-occurrence composition in three metrics. This may further show that the adjacent reading sequence, either directed or undirected, may play an important role, and the co-occurrence graph ignores such information. 

\begin{figure*}[htbp]
    \centering
    \includegraphics[width=0.98\textwidth]{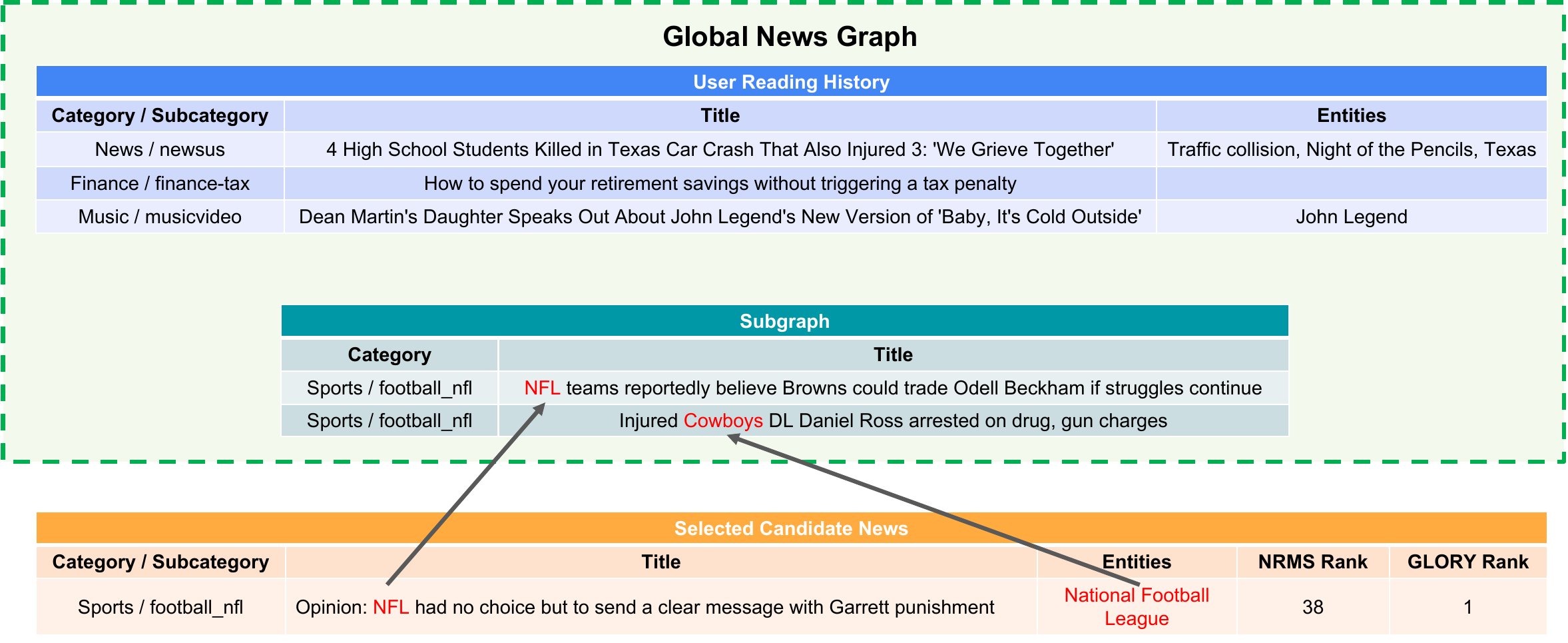}
    \caption{Case study based on a sampled impression. The top table displays the complete reading history of a sampled user, which includes three news articles. The middle table showcases relevant news articles selected from the subgraph, which is extracted from the global news graph based on the nodes of the news articles in the user's reading history. The bottom table presents the candidate news articles chosen by the user. The main and subcategories are only for reference. The red font and gray arrows indicate that the content in the candidate news articles can be found to have similar counterparts in the subgraph.}
    \label{fig: case}
\end{figure*}

\subsubsection{Global Entity Graph}
Additionally, for the global entity graph in Sec ~\ref{sec: gentity}, we primarily study the connection methods based on two aspects: intra-news entity connections and inter-news entity connections. In this case, we use the default directed sequential composition for the global news graph. Here are four different global entity graph connection methods: 1)  \textbf{Intra-news \& Inter-news Directed}: connects all entities within each news article, and entities between different news articles are connected in a directed manner; 2)  \textbf{Intra-news \& Inter-news Undirected}: connects all entities within each news article, and entities between different news articles are connected in an undirected manner; 3)  \textbf{Inter-news Directed}: only entities between different news articles are connected in a directed manner; 4) \textbf{Inter-news Undirected}: only entities between different news articles are connected in an undirected manner. The performance on MIND-small is shown in Tab.~\ref{tab: gentity}.

Based on the table, it appears that the model performs better without intra-news relations. One explanation could be that entities that frequently appear together in the same news article are often closely related in the knowledge graph. Therefore, intra-news relations may primarily add semantically related neighboring entities, while without them, the model can capture more diverse and global associations. Moreover, the use of inter-news undirected relations can help entities with fewer connections to have more neighboring entities, thus improving their representation in the model.

\subsection{Recommendation Diversity}

In this section, we evaluate the recommendation diversity of GLORY and NRMS using the intra-list average distance (ILAD@N) and the intra-list minimum distance (ILMD@N) as metrics, with $N={5,10}$. These metrics are commonly used to measure the average and minimum distances between recommendation items \cite{diversity1, diversity2}. Following \cite{end-to-end}, we use average GloVe word embeddings as semantic embeddings, instead of news embeddings learned by the model because they cannot be compared across models. 
The results obtained on MIND-small are presented in Tab.~\ref{tab: diversity}, which clearly demonstrates the effectiveness of GLORY in improving recommendation diversity. The NRMS model only recommends news based on local features learned from the historical news of the current user, while GLORY leverages global information beyond semantics to provide a wider range of content diversity. This makes GLORY particularly effective in enhancing recommendation diversity. By taking a more holistic approach to content analysis, GLORY is able to provide a richer and more varied set of recommendations, leading to a better user experience.

\begin{table}[htbp]
    \centering
    \caption{Comparison of recommendation diversity on MIND-small.}
    \begin{tabular}{c|c c c c}
    \toprule
        Model & ILAD@5 & ILMD@5 & ILAD@10 & ILMD@10 \\
        \hline
        NRMS & 0.3189 & 0.1913 & 0.3183 & 0.1486 \\
        GLORY & \textbf{0.3226} & \textbf{0.2167} & \textbf{0.3329} & \textbf{0.1848} \\
        \bottomrule
    \end{tabular}
    \label{tab: diversity}
\end{table}

\section{Case Study}

Finally, we conduct a case study to examine the superior performance of GLORY specifically. We compare GLORY with NRMS because using the same structure for the user encoder allows us to showcase the excellence of our news modeling better. In Fig.~\ref{fig: case}, we present a random sample from the MIND-small test dataset, where the corresponding user has three history news varies from categories news-us to finance, while the candidate news the user read is about National Football League (NFL). Merely analyzing the user's interests semantically, NRMS is insufficient to uncover the association between historical news and the selected candidate news. In contrast, our model, GLORY, is able to identify several neighbor news articles related to NFL in the global news graph. Furthermore, GLORY recognizes that the entity "NFL" in the candidate news article was highly relevant to "Cowboys", which is an NFL team. Consequently, GLORY ranks the selected candidate news article in the first position, while NRMS ranks it at the 38th position.

\section{Limitations}
Furthermore, this paper explores the limitations of our approach. First, our approach faces an efficiency issue during training as GLORY needs to acquire and process global information from both the global news graph and global entity graph for each sample. Consequently, this leads to increased memory and time requirements during training, compared to using only local information. Secondly, our approach trains and validates using two public news datasets containing click data from limited time periods, this restricts us to use a static global graph and precludes testing on dynamically changing real-world data.

\section{Conclusion and Future Work}
This paper presents GLORY, a novel news recommendation system that utilizes global graphs to improve news modeling. We propose the global news graph to capture hidden behavior and patterns of all users' reading history. This approach addresses the limitations of relying solely on semantic news modeling. Additionally, we employ the global entity graph to uncover deeply hidden associations between candidate news articles and historical news articles. Our approach enhances news recommendations by leveraging both global and local information.
We use historical news aggregator and candidate news aggregator to fuse local embeddings and global embeddings, enriching the implied content of news representations. Finally, we incorporate a concise multi-head self-attention-based user encoder. Through a series of experiments on two datasets, we reveal that our model significantly outperforms all other state-of-the-art models and validate the contributions of different components to the model. As for future work, we will continue exploring dynamic global graphs that consider the freshness of news items and user behaviors, making them suitable for real-time online recommendation systems.

\begin{acks}
This research is partially supported by JSPS KAKENHI Grant Numbers JP21K21280 and JP23H03408. It is also partially supported by the Initiative on Recommendation Program for Young Researchers and Woman Researchers, Information Technology Center, The University of Tokyo. Additionally, the research is also supported by Foundation for Computational Science (FOCUS) Establishing Supercomputing Center of Excellence, Japan through RIKEN Center for Computational Science (R-CCS). This research was conducted with the financial support of Science Foundation Ireland 12/RC/2289\_P2 at Insight the SFI Research Centre for Data Analytics at University College Dublin. We thank the anonymous reviewers for their insightful comments.
\end{acks}
\balance

\bibliographystyle{ACM-Reference-Format}
\bibliography{main}

\appendix

\end{document}